\documentclass{ws-jtca}
\usepackage[super]{cite}
\usepackage{lineno}
\usepackage{color}
\usepackage{subfigure}

\begin{document}

\markboth{H. Tu et al.}{Chebyshev-Tau spectral method to solve the parabolic equation model}

\catchline{}{}{}{}{}

\title{Applying the Chebyshev-Tau spectral method to solve the parabolic equation model of wide-angle rational approximation in ocean acoustics}

\author{Houwang Tu, Yongxian Wang, Xian Ma and Xunjiang Zhu}
\address{College of Meteorology and Oceanography, National University of Defense Technology,\\Changsha, 410073, China\\
\email{tuhouwang96@163.com}\ \email{yxwang@nudt.edu.cn} }

\maketitle

\begin{abstract}
Solving an acoustic wave equation using a parabolic approximation is a popular approach for many existing ocean acoustic models. Commonly used parabolic equation (PE) model programs, such as the range-dependent acoustic model (RAM), are discretized by the finite difference method (FDM). Considering the idea and theory of the wide-angle rational approximation, a discrete PE model using the Chebyshev spectral method (CSM) is derived, and the code is developed. This method is currently suitable only for range-independent waveguides. Taking three ideal fluid waveguides as examples, the correctness of using the CSM discrete PE model in solving the underwater acoustic propagation problem is verified. The test results show that compared with the RAM, the method proposed in this paper can achieve higher accuracy in computational underwater acoustics and requires fewer discrete grid points. After optimization, this method is more advantageous than the FDM in terms of speed. Thus, the CSM provides high-precision reference standards for benchmark examples of the range-independent PE model.
\end{abstract}


\section{Introduction}

The ocean contains an abundance of energy, minerals and biological resources. The urgent requirements of marine research and development have posed new challenges for the detection, identification, positioning and communication of underwater targets. At present, sound waves are the main means for remotely transmitting information underwater; therefore, it is of great practical significance to thoroughly study and understand the laws of underwater acoustic propagation. As a mathematical expression of acoustic physical properties, a numerical acoustic field can describe the physical laws of ocean acoustic propagation with simple and clear numerical solutions. Commonly used computational ocean acoustic theories include the parabolic equation (PE) model, normal modes, the wavenumber integration method and the ray model \cite{Finn2011}. 

Among these models, the PE model has the advantage of being fast and flexible when solving range-dependent acoustic propagation problems. In the 1970s, Hardin and Tappert \cite{Hardin1973,Tappert1977} introduced the PE method in the field of underwater acoustics for the first time and approximated the Helmholtz equation as a two-dimensional equation that was related only to the range and depth and was independent of the azimuth. In the 1980s, Davis et al. derived a generalized PE model using the operator method \cite{Davis1982}; the derivation based on a series expansion of the square-root operator $Q$ enables the formulation of better PE approximations with a wide-angle capability. Greene \cite{Greene1984} and Claerbout \cite{Claerbout1985} selected different coefficients and derived their respective wide-angle PE models. Accordingly, interest in PE techniques has steadily grown within the ocean acoustic modeling community. Based on the idea of parabolic approximation, many parabolic model schemes have been proposed \cite{Lee1981,Mcdaniel1982,Botseas1983,Lee1988}. More types of PE models can be found in the literature \cite{Gilbert1993,Salomons1998,Gilbert2016}. Bamberger et al. presented a generalized very-wide-angle PE based on a Pad\'e series expansion \cite{Bamberger1988}. Collins was the first to implement the wide-angle PE numerical solution based on the high-order Pad\'e approximation \cite{Collins1989,Collins1991} and expanded the propagation angle to nearly 90°. This process solved many practical problems, such as the self-starter to obtain an initial condition \cite{Collins1992}, ``split-step'' high-order Pad\'e series approximation \cite{Collins1993a}, energy loss problem caused by a step approximation \cite{Collins1993b,Siegmann1999} and treatment of an inclined seafloor boundary \cite{Collins1990}. Then, the classic underwater range-dependent acoustic model (RAM) program was developed, and the depth operator was discretized using the finite difference method (FDM). This approach of replacing the depth operator with a tridiagonal matrix can address piecewise continuous depth variations in acoustic parameters \cite{Canuto2006}. After discretizing in the depth direction, the numerical solution involves repeatedly solving tridiagonal systems of equations. However, although the FDM is the main approach for numerical simulations in computational underwater acoustics (RAM, FOR3D, FFP, KRAKEN, and Bellhop), it still has many shortcomings and deficiencies, such as difficulty in constructing high-precision difference schemes; furthermore, the convergence and stability of complex numerical schemes are difficult to verify mathematically. In addition, the equidistant grid used by the FDM can cause information that is smaller than the grid size to be ignored. Thus, to satisfy the rapidly growing requirements of various underwater acoustic applications, the development of new discrete methods with high accuracy and speed has important scientific and application value.

In scientific computing and numerical simulations in engineering, the spectral method (SM), FDM and finite element method are the three major discrete numerical methods \cite{Peyret1986,Xiang2000}. Because of its high accuracy, the SM is frequently used in various mathematical and physical problems, such as computational fluid dynamics \cite{Canuto2006}, chemical measurements \cite{Muravskaya2019} and electricity \cite{Ammi2017}. The SM originates from the method of weighted residuals. It uses orthogonal polynomials (triangular polynomials, Chebyshev polynomials, Legendre polynomials) as the basis functions and applies finite-term series to approximate the variables to be solved. The greatest advantage of the SM is that it exhibits exponential convergence; i.e., when the solution of the original equation is sufficiently smooth, the approximate solution obtained by the SM will quickly converge to the exact solution. In recent decades, the SM has been vigorously developed and has become an important tool for solving differential equations \cite{Trefethen1996}. Some researchers have begun to apply the SM to solve acoustic problems. Wise \cite{Wise2019} presented an arbitrary acoustic source and sensor distributions using the Fourier collocation method. Wang et al. \cite{Wang2006} presented an improved expansion scheme for the acoustic wave propagator. Evans \cite{Evans2016} proposed a Legendre-Galerkin technique for differential eigenvalue problems with complex and discontinuous coefficients in underwater acoustics. Subsequently, Evans \cite{Evans2018} studied the Legendre-Galerkin SM to construct atmospheric acoustic normal modes. Most recently, Tu et al. \cite{Tu2021} implemented a Chebyshev-Tau SM to solve acoustic normal modes with a stratified ocean.

In applying the SM to solve the PE model, Tu et al. \cite{Tu2020} presented the standard PE model using the Chebyshev spectral method (CSM) to process a single layer of a body of water with constant density and no attenuation. To date, however, no research has been conducted on the use of the SM to solve the PE model of wide-angle rational approximation. This paper introduces the CSM into the solution of the PE model in underwater acoustics, presents three examples, and finally analyzes and compares the advantages and shortcomings of the proposed CSM-based PE model in terms of its computational accuracy and speed.

\section{Derivation of the PE model}
Consider a cylindrical coordinate system in which the sound source is a simple harmonic point source and the marine environment is a cylindrically symmetric two-dimensional medium of acoustic propagation. The Helmholtz equation can be written as \cite{Finn2011}:
	\begin{equation}
		\label{eq:1}
		\left[\frac{\partial^2}{\partial r^2}+\frac{1}{r} \frac{\partial}{\partial r}+\rho \frac{\partial}{\partial z} \left(\frac{1}{\rho}\frac{\partial}{\partial z}\right)+k_0^2\tilde{n}^2 \right]\tilde{P}=0,
	\end{equation}
where $\rho$ is the density, $c_0$ is the reference sound speed, $f$ is the frequency of the sound source, $k_0=2\pi f/c_0$ is the reference wavenumber, $\tilde{n}=c_0/c$ is the refractive index, and $\tilde{P}$ is the acoustic pressure in the frequency domain. For long-range acoustic propagation, sound waves are generally approximated as cylindrical waves. According to the attenuation law of cylindrical waves, the energy amplitude of the sound wave is proportional to $\sqrt{r}$. To eliminate the extension term, the following coordinate transformation is introduced for the acoustic pressure in Eq.~\eqref{eq:1}:
	\begin{equation}
		\label{eq:2}
		p=\sqrt{r}\tilde{P}.
	\end{equation}
By substituting Eq.~\eqref{eq:2} into Eq.~\eqref{eq:1}, we obtain:
	\begin{equation}
		\label{eq:3}
		\frac{\partial^2p}{\partial r^2}+\rho \frac{\partial}{\partial z} \left(\frac{1}{\rho}\frac{\partial p}{\partial z}\right)+k_0^2\tilde{n}^2p+\frac{p}{4r^2}=0.
	\end{equation}
By applying the ``far-field'' approximation $k_0 r\gg 1 $ to Eq.~\eqref{eq:3}, we know:
	\begin{equation}
		\label{eq:4}
    	k_0^2\tilde{n}^2p\gg\frac{p}{4r^2}.
	\end{equation}
After discarding the small term $\frac{p}{4r^2}$, we can rewrite Eq.~\eqref{eq:4} as follows:
	\begin{equation}
		\label{eq:5}
		\frac{\partial^2p}{\partial r^2}+\rho \frac{\partial}{\partial z} \left(\frac{1}{\rho}\frac{\partial p}{\partial z}\right)+k_0^2\tilde{n}^2p=0.
	\end{equation}
When considering attenuation, let the wavenumber $k=(1+i\eta\beta)\omega/c(z)$, where $c(z)$ is the sound speed, $\omega=2\pi f$ is the circular frequency, $\beta$ is the attenuation in dB/wavelength, $i$ indicates an imaginary number, and $\eta=(40\pi \log_{10}e)^{-1}$. We introduce the operator decomposition method proposed by Lee et al. \cite{Lee1981} to factorize Eq.~\eqref{eq:5} and rewrite it in the form of outwardly and inwardly propagating waves:
	\begin{equation}
		\label{eq:6}
		\left(\frac{\partial}{\partial r}-ik_0\sqrt{1+\mathcal{X}}\right)\left(\frac{\partial}{\partial r}+ik_0\sqrt{1+\mathcal{X}}\right) p +\left[\frac{\partial}{\partial r},ik_0 \sqrt{1+\mathcal{X}}\right]p = 0,
	\end{equation}
where the square brackets indicate the commutator of the operators. The depth operator $\mathcal{X}$ is expressed as:
	\begin{equation}
		\label{eq:7}
		\mathcal{X}=k_0^{-2}\left[\rho(z) \frac{\partial}{\partial z}
		\left(
		\frac{1}{\rho (z)}\frac{\partial }{\partial z}
		\right)+k^2-k_0^2\right].
	\end{equation}
When we ignore the horizontal variation in the medium parameter, the derivative terms of $\left[\frac{\partial}{\partial r},ik_0 \sqrt{1+\mathcal{X}}\right]$ in the above formula can be exchanged; therefore, the simplified form of Eq.~\eqref{eq:6} is as follows:
	\begin{equation}
		\label{eq:8}
		\left(\frac{\partial}{\partial r}-ik_0\sqrt{1+\mathcal{X}}\right)\left(\frac{\partial}{\partial r}+ik_0\sqrt{1+\mathcal{X}}\right) p = 0.
	\end{equation}
Among the two terms of Eq.~\eqref{eq:8}, the first term represents waves propagating outwards, while the second term represents waves propagating inwards, and inward propagation is negligible. Then, we obtain the PE in the following form:	
	\begin{equation}
		\label{eq:9}
		\frac{\partial p}{\partial r}=ik_0\sqrt{1+\mathcal{X}}p.
	\end{equation}
According to the method of solving ordinary differential equations, the step solution of the equation can be obtained as:
	\begin{equation}
		\label{eq:10}
		p(r+\Delta r,z)=\exp{\left(ik_0\Delta r\sqrt{1+\mathcal{X}}\right)}p(r,z),
	\end{equation}
where $\Delta r$ is the step size in the horizontal direction. Collins used the Pad\'e series expansion method \cite{Salomons1998}, where an $n$-term rational function is used to approximate the exponential function in Eq.~\eqref{eq:10}:
	\begin{equation}
		\label{eq:11}
		p(r+\Delta r,z)=\exp{(ik_0\Delta r)}\prod_{j=1}^{n}\frac{1+\alpha_{j,n}\mathcal{X}}{1+\beta_{j,n}\mathcal{X}}p(r,z),
	\end{equation}
$n$ is the number of items used for the rational approximation; the choice will also affect the accuracy of the approximation. More precisely, it is determined by the user according to the complexity of the research environment and the characteristics of the sound source. The complex coefficients $\alpha_{j,n}$ and $\beta_{j,n}$ must satisfy the stability, convergence and accuracy requirements.

To numerically solve Eq.~\eqref{eq:11}, the depth operator $\mathcal{X}$ must be discretized. Traditionally, the FDM is often used to discretize and form a set of tridiagonal matrix algebraic equations. In this study, the CSM is used to solve the PE model.
	
\section{CSM for the PE model}
\subsection{Chebyshev spectral method}
The SM is derived from the method of weighted residuals, which is based on the expansion and summation of a finite series to approximate the unknown function to be solved. Any continuous and sufficiently smooth unknown function $u(x)$ is expanded by a set of smooth bases and weighted sums. Essentially, this set of bases forms a function space, and the so-called expansion can also be understood as a projection \cite{Boyd2001}. The SM using Chebyshev orthogonal polynomials as the basis functions is the CSM, in which the basis functions $T_k (x)$ are defined as:
    \begin{equation}
        \begin{split}
        \label{eq:12}
            T_{0}(x)&=1 ; \quad T_{1}(x)=x ; \quad T_{2}(x)=2 x^{2}-1\\
            T_{k+1}(x)&=2 x T_{k}(x)-T_{k-1}(x),\quad x \in [-1,1],
            \quad k = 2, 3, 4, \cdots.
        \end{split}
    \end{equation}
Any smooth and differentiable unknown function $u(x)$ can be expanded by the infinite basis functions and approximated by the finite sum of the first $(N+1)$ terms:
    \begin{equation}
    \label{eq:13}
        u(x)=\sum_{k=0}^{\infty}\hat{u}_k T_k(x) \approx \sum_{k=0}^{N}\hat{u}_k T_k(x).
    \end{equation}
The expansion coefficients can be obtained by the following formula:
    \begin{equation}
    \label{eq:14}
        \hat{u}_k = \frac{2}{\pi c_k} \int_{-1}^{1} \frac{u(x)T_k(x)}{\sqrt{1-x^2}} \mathrm{d} x,
        \quad c_k=\begin{cases}
            2,  & k=0\\
            1,  & k>0
        \end{cases}.
    \end{equation}
Eqs.~\eqref{eq:13} and \eqref{eq:14} are the so-called backward Chebyshev transform and forward Chebyshev transform, respectively.

The integral in Eq.~\eqref{eq:14} is usually calculated numerically. The Gaussian quadrature method is generally considered to have high accuracy. The Chebyshev-Gauss-Lobatto points are often used in the Gauss-Chebyshev-Lobatto quadrature method \cite{Jieshen2011}:
    \begin{equation}
    \label{eq:15}
        x_j=\cos{\left(\frac{j\pi}{N}\right)},
        \quad j=0,1,2,\dots,N,
    \end{equation}
where $(N+1)$ is the number of discrete points, which is equal to the truncated order in the CSM. When $(N+1)$ Chebyshev-Gauss-Lobatto points are used, the expansion coefficient $\hat{u}_k$ in Eq.~\eqref{eq:15} can be approximately obtained as:
    \begin{equation}
    \label{eq:16}
    \begin{split}
        \hat{u}_k \approx \frac{1}{d_k}\sum_{j=0}^{N}u(x_j)T_k(x_j)\omega_j,
        \quad k=0,1,2,\dots,N \\
        \omega_j=\begin{cases}
        \frac{\pi}{2N},\quad j=0,N  \\
        \frac{\pi}{N}, \quad \text{otherwise}
     \end{cases},\quad
        d_k=\begin{cases}
        \pi,\quad k=0,N  \\
        \frac{\pi}{2},\quad \text{otherwise}
     \end{cases}.
    \end{split}
    \end{equation}
Similarly, the first derivative of $u(x)$ can also be expanded as:
    \begin{equation}
    \label{eq:17}
        u'(x) = \sum_{k=0}^{\infty}\hat{u}'_k T_k(x) \approx \sum_{k=0}^{N}\hat{u}'_k T_k(x).
    \end{equation}
Considering the first derivative, due to the relationship between the Chebyshev polynomial and its derivative, the following relationship between $\hat{u}'_k$ and $\hat{u}_k$ can be expressed as follows:
    \begin{equation}
    \label{eq:18}
        \hat{u}'_k \approx \frac{2}{c_k}
            \sum_{
            \substack{p=k+1,\\ 
                p+k=\text{odd}
                }}^{N} p \hat{u}_p,\quad c_k=\begin{cases}
        2,\quad k=0  \\
        1,\quad \text{otherwise}
        \end{cases}.
    \end{equation}
In Eq.~\eqref{eq:18}, the expansion coefficients can be written as a column vector, and the algebraic relationship between $\hat{u}'_k$ and $\hat{u}_k$ can be written as a matrix. Let the relationship matrix be $\mathbf{D}$. Then, Eq.~\eqref{eq:18} can be expressed as:
    \begin{equation}
    \label{eq:19}
        \mathbf{\hat{u}}' \approx \mathbf{D}\mathbf{\hat{u}},\quad
        \mathbf{\hat{u}}^{(m)} \approx \mathbf{D}^m \mathbf{\hat{u}},
    \end{equation}
where $m$ is an arbitrary natural number and $\mathbf{D}$ is a square matrix \cite{Tu2021}.

The product $w(x)$ of two functions $u(x)$ and $v(x)$ is called a convolution term in the SM. The spectral coefficients of the $w(x)$ have a relationship with the respective spectral coefficients \cite{Mason2003}:
    \begin{equation}
    \label{eq:20}
        \hat{w}_k \approx 
            \frac{1}{2} \sum_{m+n=k}^{N} \hat{u}_m\hat{v}_n +
            \frac{1}{2} \sum_{|m-n|=k}^{N} \hat{u}_m\hat{v}_n.
    \end{equation}
The algebraic relationship between $\hat{w}_k$ and $\hat{u}_k$ can also be written as a matrix, and the relationship matrix can be called $\mathbf{C}_v$ \cite{Tu2021}; thus, Eq.~\eqref{eq:20} becomes:
    \begin{equation}
    \label{eq:21}
        \mathbf{\widehat{w}} \approx \mathbf{C}_v \mathbf{\hat{u}}.
    \end{equation}

\subsection{Discrete PE model using the CSM}
In the following section, we use the CSM to numerically discretize the operator $\mathcal{X}$ in Eq.~\eqref{eq:7} and $p$ in Eq.~\eqref{eq:11}. First, a linear transformation $x=1-2z/H$ is used to transform the domain of the original problem from $z\in [0,H]$ to $x\in [-1,1]$. Then, the depth operator $\mathcal{X}$ becomes:
	\begin{equation}
		\label{eq:22}
		\mathcal{X}=k_0^{-2}\left[\frac{4}{H^2}\rho(x) \frac{\partial}{\partial x}
		\left(
		\frac{1}{\rho(x)}\frac{\partial }{\partial x}
		\right)+k^2-k_0^2\right].
	\end{equation}
Thus, Eq.~\eqref{eq:11} becomes:
	\begin{equation}
		\label{eq:23}
		p(r+\Delta r,x)=\exp{(ik_0\Delta r)}\prod_{j=1}^{n}\frac{1+\alpha_{j,n}\mathcal{X}}{1+\beta_{j,n}\mathcal{X}}p(r,x).
	\end{equation}
When applying the CSM, it is necessary to consider the $N$-term truncation of the infinite term expansion; the problem of solving the ordinary differential problem for an unknown function $p(r+\Delta r,x)$ is transformed into an algebraic problem with $(N+1)$ unknowns. Among the available equation-constructed methods, common methods include the Tau, Galerkin and collocation methods. In this article, we mainly consider the Tau method. The Chebyshev-Tau SM aims to obtain the solution of the weak form of Eq.~\eqref{eq:23} (see Eq. (3.3.15) of Ref.~\cite{Canuto2006} for further details about the weak form):
    \begin{equation}
    \label{eq:24}
        \int_{-1}^1 \left[p(r+\Delta r,x)-\exp{(ik_0\Delta r)}\prod_{j=1}^{n}\frac{1+\alpha_{j,n}\mathcal{X}}{1+\beta_{j,n}\mathcal{X}}p(r,x)  \right]\frac{T_k(x)}{\sqrt{1-x^2}} \mathrm{d} x=0,
    \end{equation}
where $k=0,1,2,\dots,(N-2)$. In each step iteration, $k$ changes from 0 to $(N-2)$ to produce $(N-1)$ linear equations (orthogonal Chebyshev polynomial bases); the boundary conditions of the sea surface and bottom will produce two linear equations, and these $(N+1)$ equations are solved simultaneously.

In the CSM, the depth operators in Eq.~\eqref{eq:22} are processed using Eqs.~\eqref{eq:19} and \eqref{eq:21}, and the discrete depth operator $\mathcal{X}$ in matrix form is obtained:
	\begin{equation}
		\label{eq:25}
		\mathbf{X}=k_0^{-2}\left(\frac{4}{H^2}\mathbf{C}_{\rho}\mathbf{D}\mathbf{C}_{1/\rho}\mathbf{D}+\mathbf{C}_{k^2}-k_0^2\mathbf{I}\right),
	\end{equation}
where $\mathbf{I}$ is the identity matrix and $\mathbf{C}_{\rho}$ and $\mathbf{C}_{1/\rho}$ are the convolution matrices of $\rho(x)$ and $1/\rho(x)$, respectively. If the seawater density is vertically uniform, then $\mathbf{C}_{\rho} \mathbf{C}_{1/\rho}=\mathbf{I}$. The forward Chebyshev transform is used to convert Eq.~\eqref{eq:11} from the original physical space to the spectral space with the $N$-term truncation approximation, and we finally obtain:
	\begin{equation}
		\label{eq:26}
		\mathbf{\hat{p}}(r+\Delta r,x)=\exp{(ik_0\Delta r)}\prod_{j=1}^{n}\frac{1+\alpha_{j,n}\mathbf{X}}{1+\beta_{j,n}\mathbf{X}}\mathbf{\hat{p}}(r,x),
	\end{equation}
where $\mathbf{\hat{p}}(r+\Delta r,x))$ and $\mathbf{\hat{p}}(r,x)$ are the column vectors that contain the $(N+1)$ Chebyshev spectral expansion coefficients of $p(r+\Delta r,x)$ and $p(r,x)$, respectively. The boundary conditions are expanded with Eq.~\eqref{eq:16} and transformed into linear equations about expanded coefficients. When the sediment is not considered, the upper and lower boundaries of the PE model are actually pressure release boundaries:
	\begin{equation}
		\tilde{P}(z=0)=\tilde{P}(z=H)=0.
	\end{equation}
From Eq.~\eqref{eq:2}, $p(z=0)=p(z=H)=0$, and the coordinate transformation is $p(x)\Big|_{x=\pm1}=0$. The boundary conditions can be expressed as:
\begin{equation}
    \sum_{k=0}^N \hat{p}_k (\pm1)^k=0.	   
\end{equation}
In actual calculations, the constraints of the boundary conditions are reflected in the matrix equations of the original problem.
The initial field $p(\Delta r)$ obtained from the self-starter \cite{Collins1992} is expanded to obtain its spectral coefficients in $\hat{p}(\Delta r)$, and the full-field $\hat{p}(r,x)$ is obtained by step iteration; then, the backward Chebyshev transform is used to transform $\hat{p}(r,x)$ into the original physical space to obtain $p(r,z)$, and Eq.~\eqref{eq:2} can be used to obtain the full-field acoustic pressure.

A previous study discussed the stability and convergence of the Chebyshev spectral approximation with the Tau method \cite{Canuto2006}. Eq.~\eqref{eq:26} is a simple boundary-value problem in each iteration. The stability and convergence of this problem have been discussed in many monographs on SMs \cite{Jieshen2011,Boyd2001}.

\section{Numerical simulation and validation}
To verify the validity of the CSM for solving the PE model, the following tests and analyses are performed with three examples. To facilitate the description, the proposed PE model program based on the CSM is abbreviated as the CSMPE. The three examples represent three types of scenarios: shallow water propagation with an analytical solution, a deep water waveguide without an analytical solution, and a case with a rough sound speed profile. Pressure release conditions are adopted at the surface and bottom of the sea. The comparison programs in the tests are the classic PE model program RAM and other programs such as KRAKEN (a classic program based on normal modes) and SCOOTER (a finite element code to compute acoustic fields in range-independent environments that is based on the direct computation of the spectral integral, and the pressure and material properties are approximated by piecewise linear elements). In addition, similar to the RAM program, the initial field of the CSMPE program in each case is also obtained by the self-starter \cite{Collins1992}. Additionally, the acoustic field traditionally displayed by the transmission loss (TL) of the acoustic pressure is defined as TL$=-20 \log_{10}(|\tilde{P}|/|\bar{P}|)$. The unit of the TL is decibel (dB), where $\bar{P}$ is the acoustic pressure at a distance of 1 m from the sound source.

\subsection{An ideal fluid waveguide with a constant sound speed}
\begin{figure}[th]
\centerline{\includegraphics[width=\linewidth]{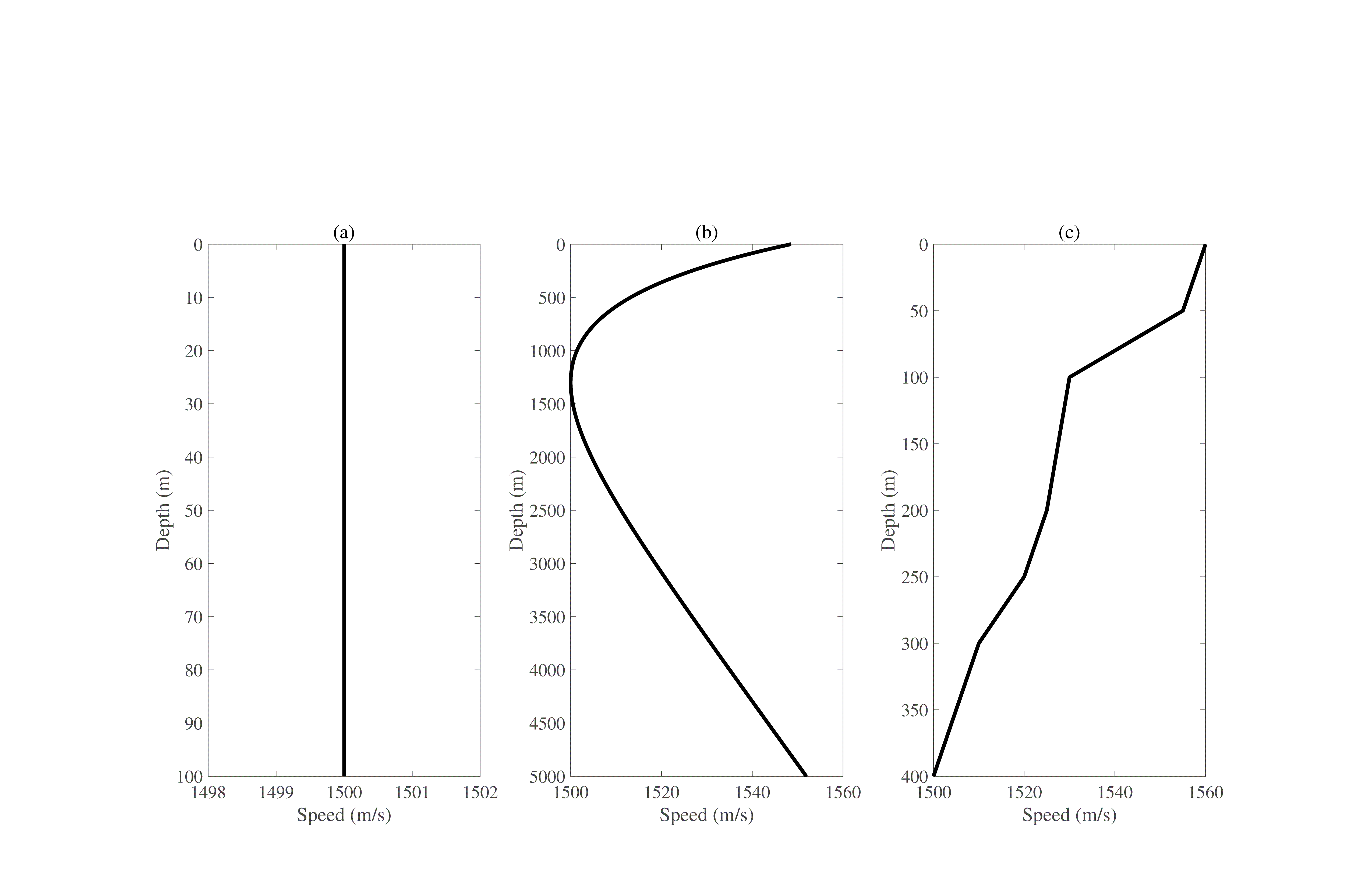}} \vspace*{8pt}
\caption{Ideal fluid marine environment with a constant sound speed in the shallow ocean (a), Munk sound speed profile in the deep ocean (b) and rough sound speed profile (c).\label{f1}}
\end{figure}

In this simple ocean waveguide model, the density of the seawater is uniform ($\rho(z)=1$ g/cm$^3$), and the sound speed underwater is unchanged ($c=c_0=1500$ m/s), as shown in Fig.~\ref{f1}(a). In this case, the depth of the sea is $H=100$ m, the sound source is located at a depth of $z_s=36$ m and the sound source frequency is $f=20$ Hz. According to the wavenumber integration method, the exact analytical solution of this ideal fluid waveguide acoustic field is \cite{Finn2011}:
	\begin{equation}
		p(r,z)=\frac{2\pi i}{H}\sum_{m=1}^{\infty}\sin{(k_{z_m}z_s)}\sin{(k_{z_m}z)}H_0^{(1)}(k_{r_m}r).
	\end{equation}
The vertical wavenumber $k_{z_m}$ and horizontal wavenumber $k_{r_m}$ are given by the following formulas:
    \begin{equation}
    	k_{z_m}=m\pi/H,\quad k_{r_m}=\sqrt{k^2-k_{z_m}^2},\quad m=1,2,\dots.
    \end{equation}

\begin{figure}[th]
\centerline{\includegraphics[width=\linewidth]{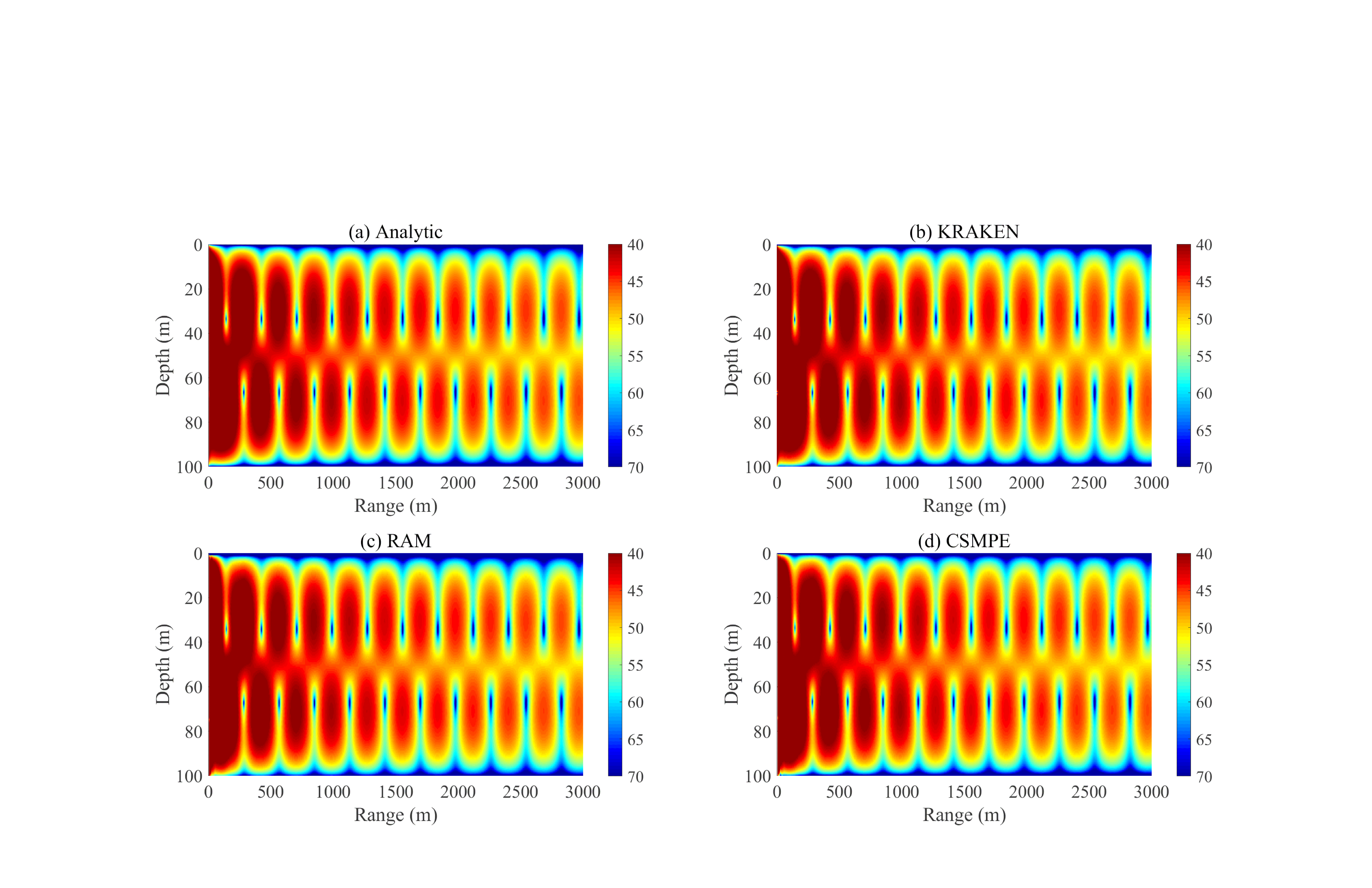}} \vspace*{8pt}
\caption{TL fields of an ideal fluid waveguide calculated using the analytical solution (a), KRAKEN (b), RAM (c), and CSMPE (d).\label{f2}}
\end{figure}

Fig.~\ref{f2} shows the TL fields of an ideal fluid waveguide calculated using the analytical solution and the KRAKEN, RAM and CSMPE programs. In the horizontal direction, $\Delta r$ in the four programs is taken as 5 m; in the vertical direction, the number of discrete grid points in KRAKEN and the RAM is taken as 200, i.e., $N_{RAM}=200$, while the CSMPE takes 25 discrete points in the vertical direction; i.e., the truncated order is $N_{CSMPE}=25$. The number of terms of the rational approximation is taken as $n=8$ (for the RAM and CSMPE). The phase velocity limit used by KRAKEN is 1500 to 2500 m/s, with a total of 2 modes. Fig.~\ref{f2} shows that the TL fields calculated by the four schemes are very similar. However, the sound fields calculated based on the PE model may have a certain degree of distortion in the near field, and the far field may have a phase error due to the introduction of the ``far-field'' approximation \cite{Finn2011}. Therefore, the correctness of the RAM and the CSMPE needs to be further compared.

\begin{figure}[th]
\centerline{\includegraphics[width=\linewidth]{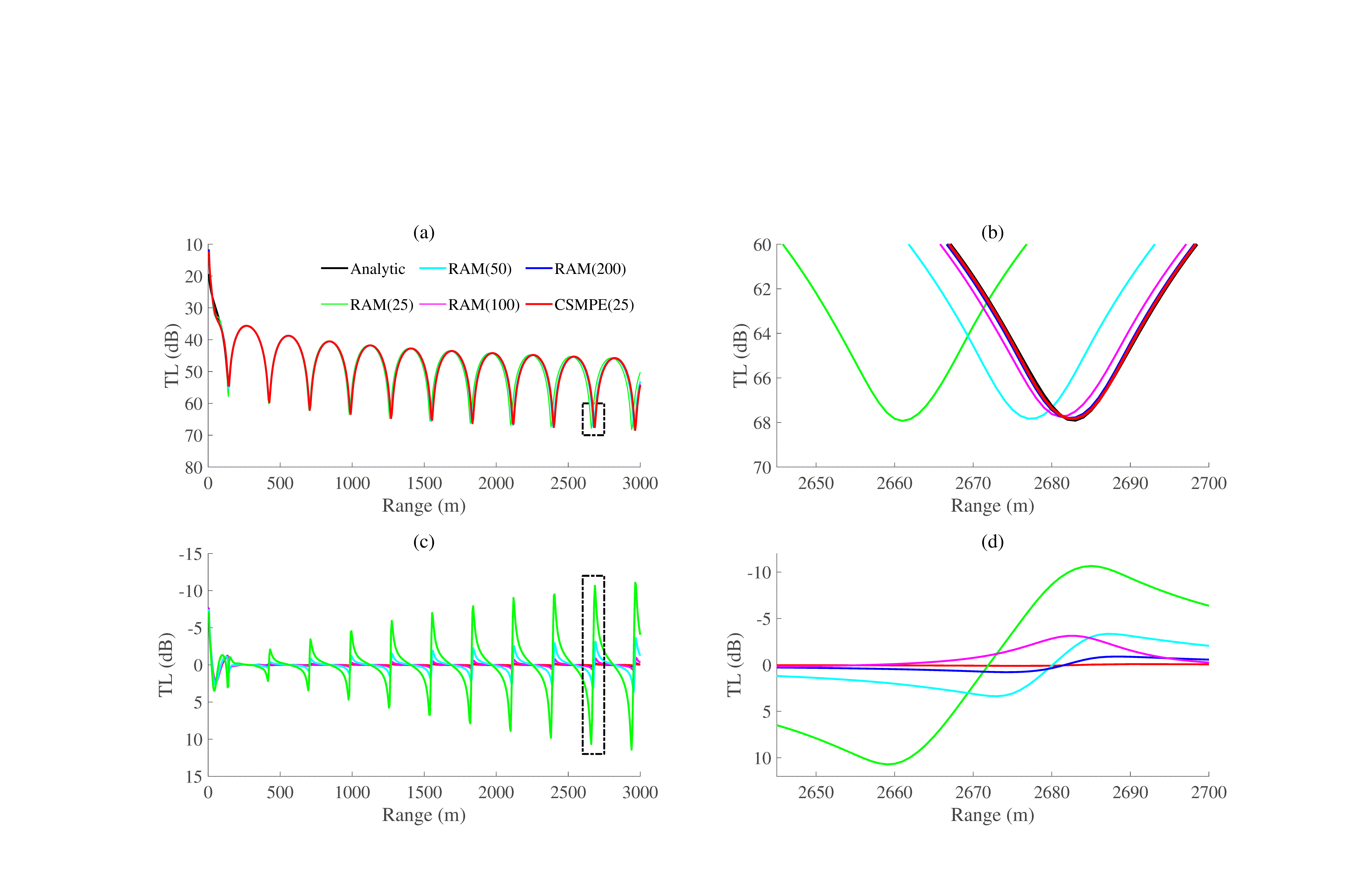}} \vspace*{8pt}
\caption{TL calculated by the programs at the depth $z_r$ (a); errors between the RAM and CSMPE results and the analytical solution (c); (b) and (d): magnified views of the dashed rectangles in (a) and (c), respectively. The numbers in the brackets after RAM and CSMPE in the legend correspond to the numbers of discrete points taken in the vertical direction.\label{f3}}
\end{figure}

To carefully compare the RAM and CSMPE results, Fig.~\ref{f3}(a) shows the TL calculated by each program at a depth of $z_r$=36 m. In the near-field area within 100 m of the source, the RAM and CSMPE results are obviously different from those of the analytical solution, but the difference between the results is relatively small. In the far field, the TL results of the four schemes are very consistent, but there are nearly inconspicuous differences in the sound shadow area. Fig.~\ref{f3}(b) shows a magnified view of the dashed rectangle in Fig.~\ref{f3}(a); when different numbers of discrete points are taken in the vertical direction, the errors between the RAM results and the analytical solution differ. More precisely, the error gradually decreases when the number of grid points increases from 25 to 200. When we take 25 discrete points in the vertical direction, the CSMPE is much more accurate than the RAM, and there is an obvious phase error between the RAM results and the analytical solution. When the number of discrete points gradually increases, the RAM phase error gradually decreases until 200 discrete points are reached. Fig.~\ref{f3}(c) shows the errors between the RAM and CSMPE results and the analytical solution, illustrating that in the near field, the results of the two PE-based programs greatly differ from those of the analytical solution (by more than 5 dB), but the error between the RAM and CSMPE results is small. In the far field, especially in the dashed rectangle, as shown in Fig.~\ref{f3}(d), the CSMPE has a smaller error than the RAM, although the latter uses more discrete points in the vertical direction ($z_r$ is exactly on the grid point, and there is no TL error caused by interpolation when $N_{RAM}$ is 25, 50, 100 or 200).

\begin{figure}[th]
\centerline{\includegraphics[width=\linewidth]{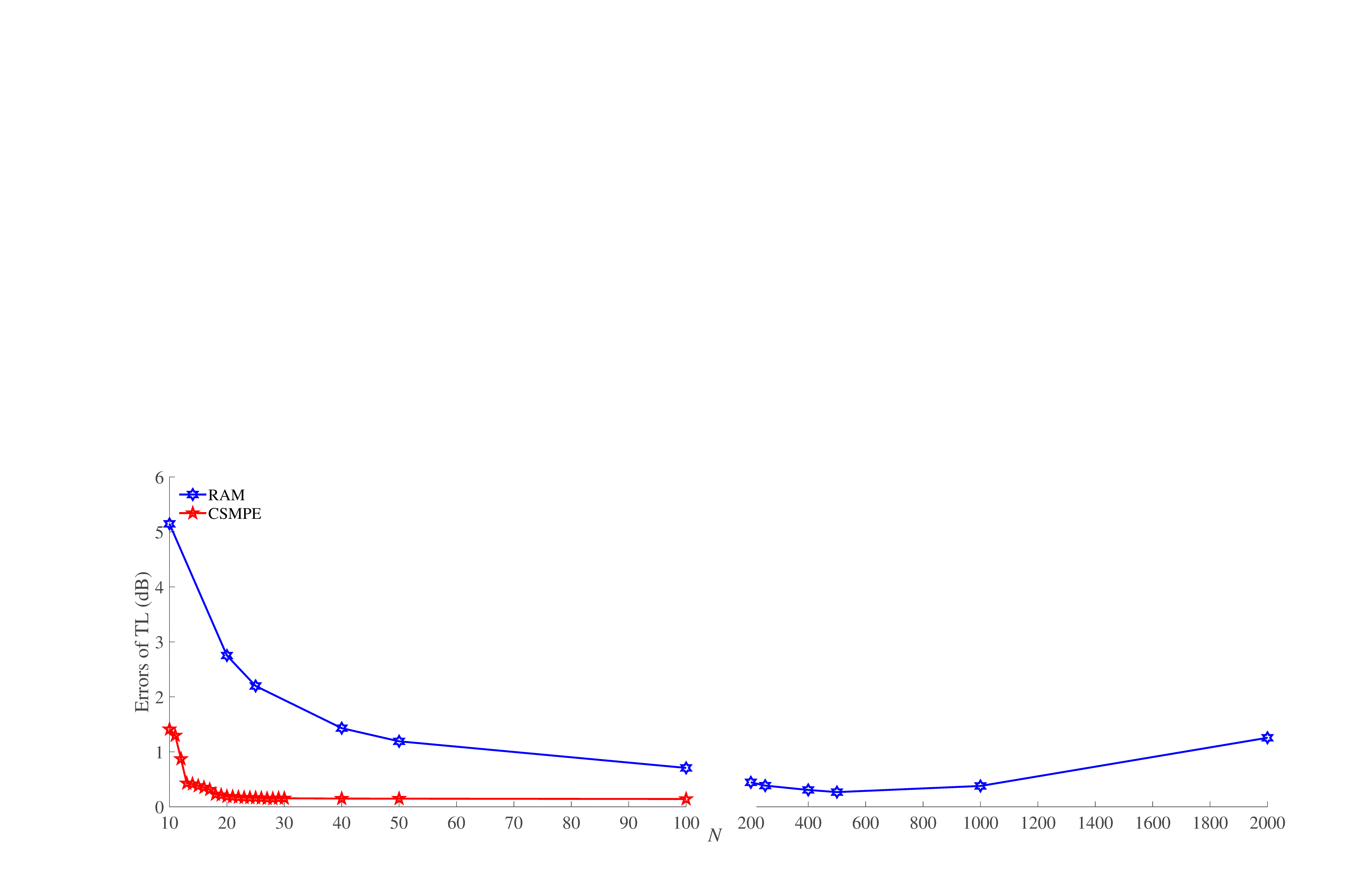}} \vspace*{8pt}
\caption{Variations in the RAM and CSMPE errors with increasing $N$.\label{f4}}
\end{figure}

Example 1 has an analytical solution, so it can be reliably used to compare the accuracy of the CSMPE and the RAM. Naturally, we define the average value of the absolute error between the acoustic pressure field and the analytical solution at discrete grid points as the error index. Fig.~\ref{f4} shows the variations in the RAM and CSMPE errors with an increase in $N$. The RAM error gradually decreases as $N_{RAM}$ increases from 20 to 500, but beyond $N_{RAM}>$500, increasing $N_{RAM}$ will make the error larger and more divergent. The CSMPE error decreases rapidly as $N_{CSMPE}$ increases from 10 to 25 and remains stable at a low level at $N_{CSMPE}>25$. In this example, the minimum error of the RAM is 0.2672 dB, while the minimum error of the CSMPE is 0.1485 dB. An important feature is that regardless of how $N_{RAM}$ is increased, the RAM error cannot be as small as the CSMPE error.

Generally, in this example, compared with the analytical solution, the results of both discrete methods are correct. Due to the ``far-field'' approximation in the PE model, the near-field results have some distortion, but in the far field, the error is extremely small. Hence, the idea of applying the CSM to the PE model is feasible. Compared with the RAM results, the CSMPE results have higher accuracy in the simple case.

\subsection{An ideal fluid waveguide with a Munk sound speed profile}
\begin{figure}[th]
\centering
\subfigure[]{\includegraphics[width=7.5cm]{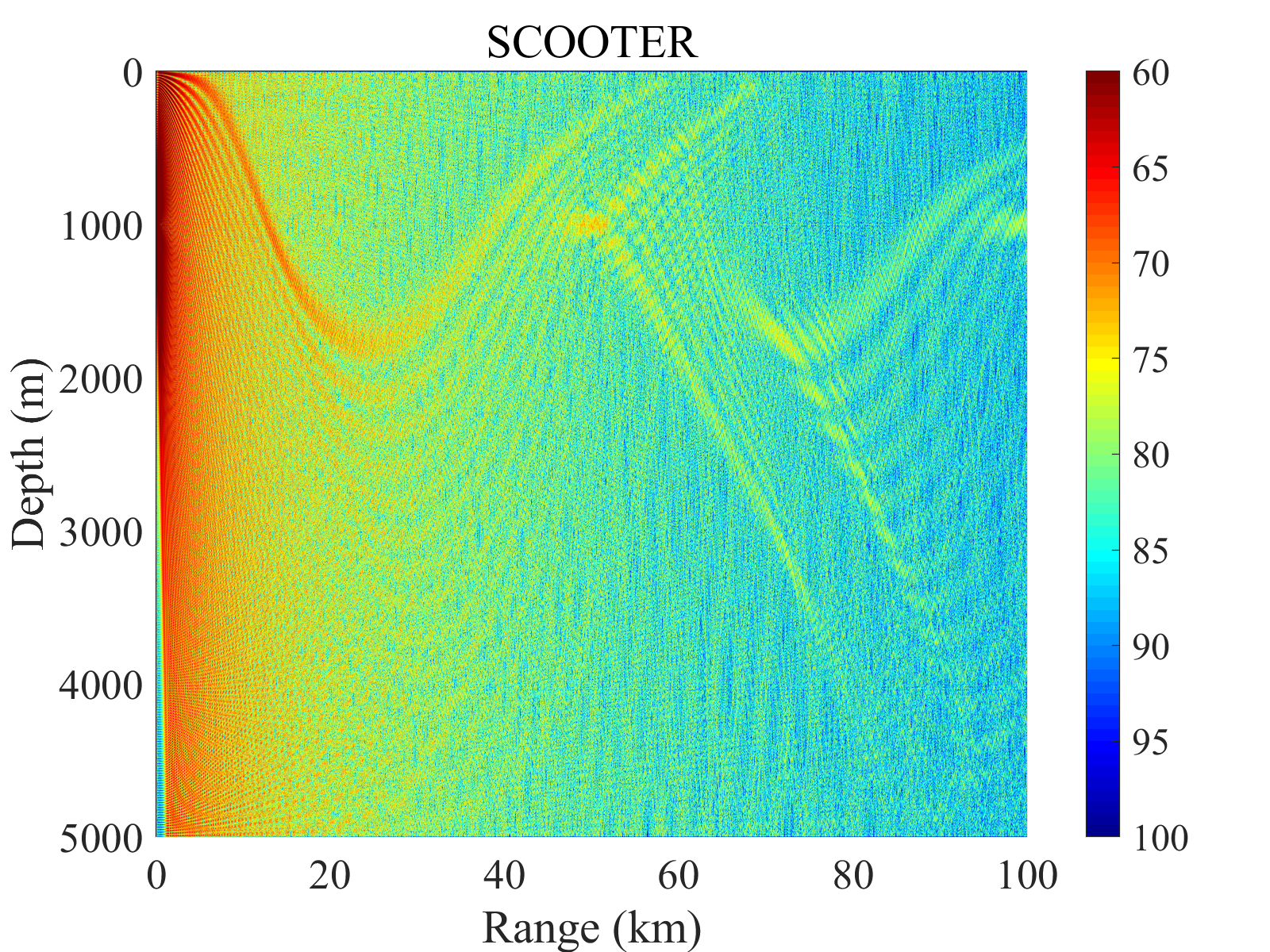}}
\subfigure[]{\includegraphics[width=7.5cm]{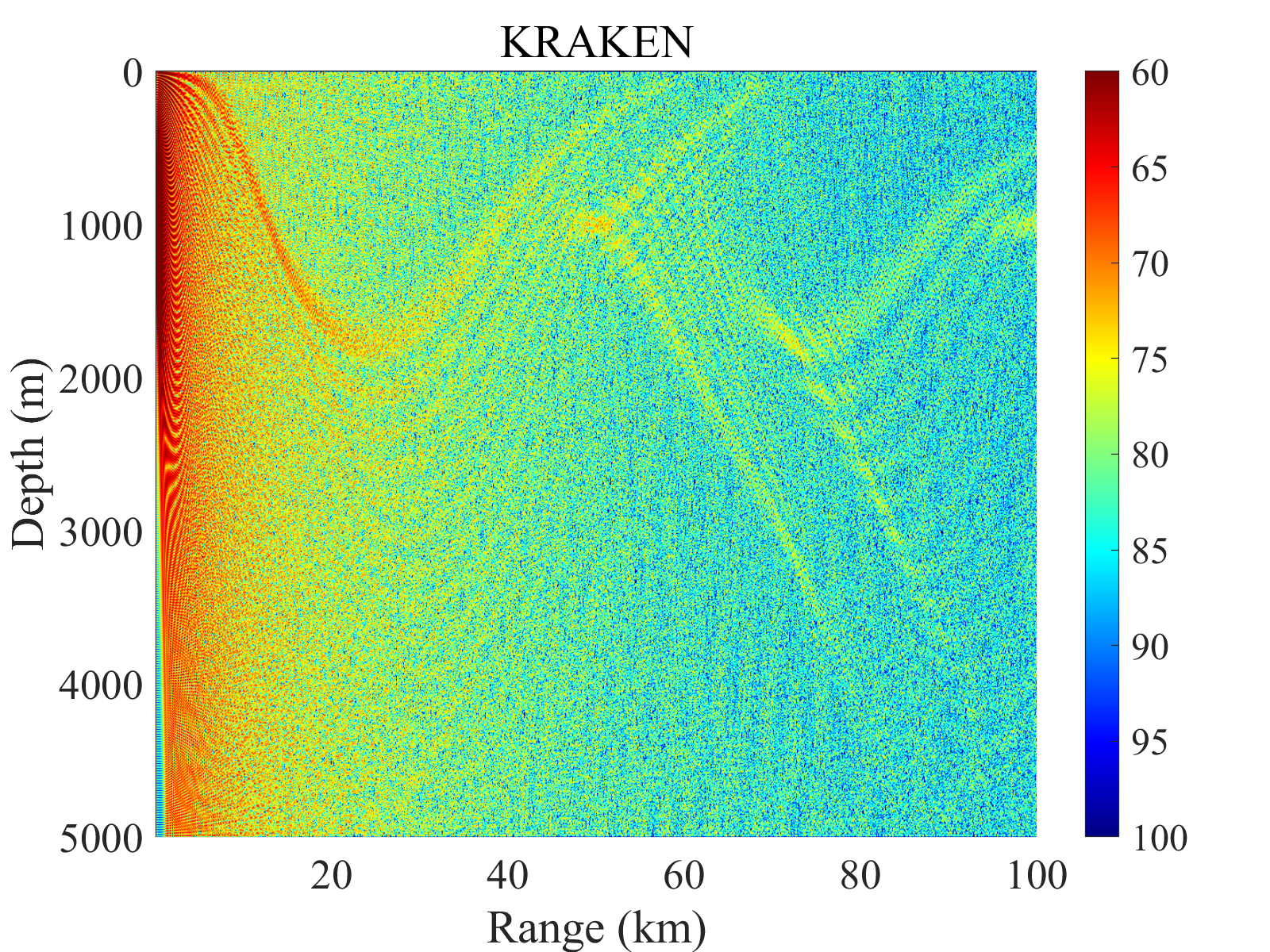}}
\subfigure[]{\includegraphics[width=7.5cm]{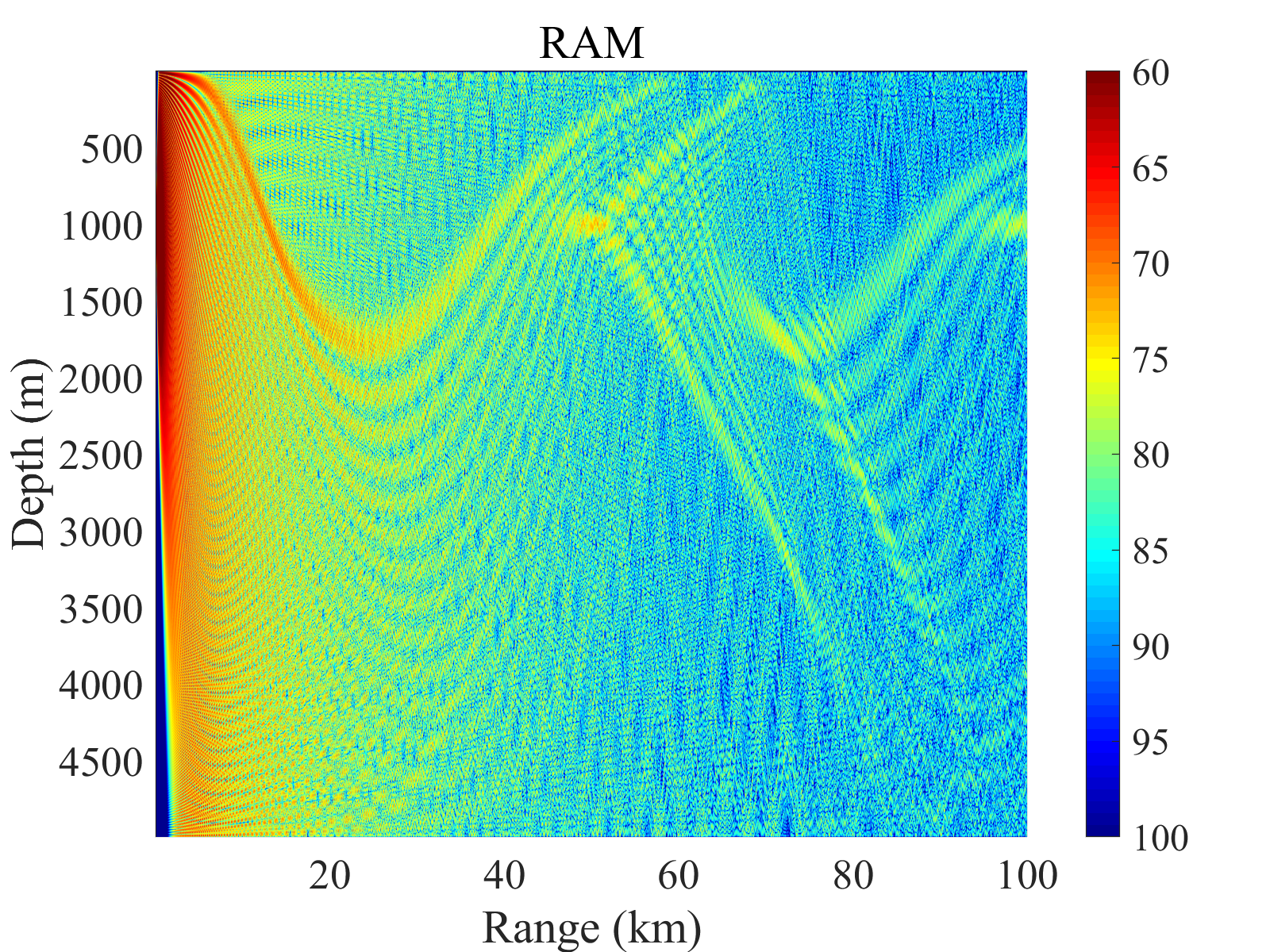}}
\subfigure[]{\includegraphics[width=7.5cm]{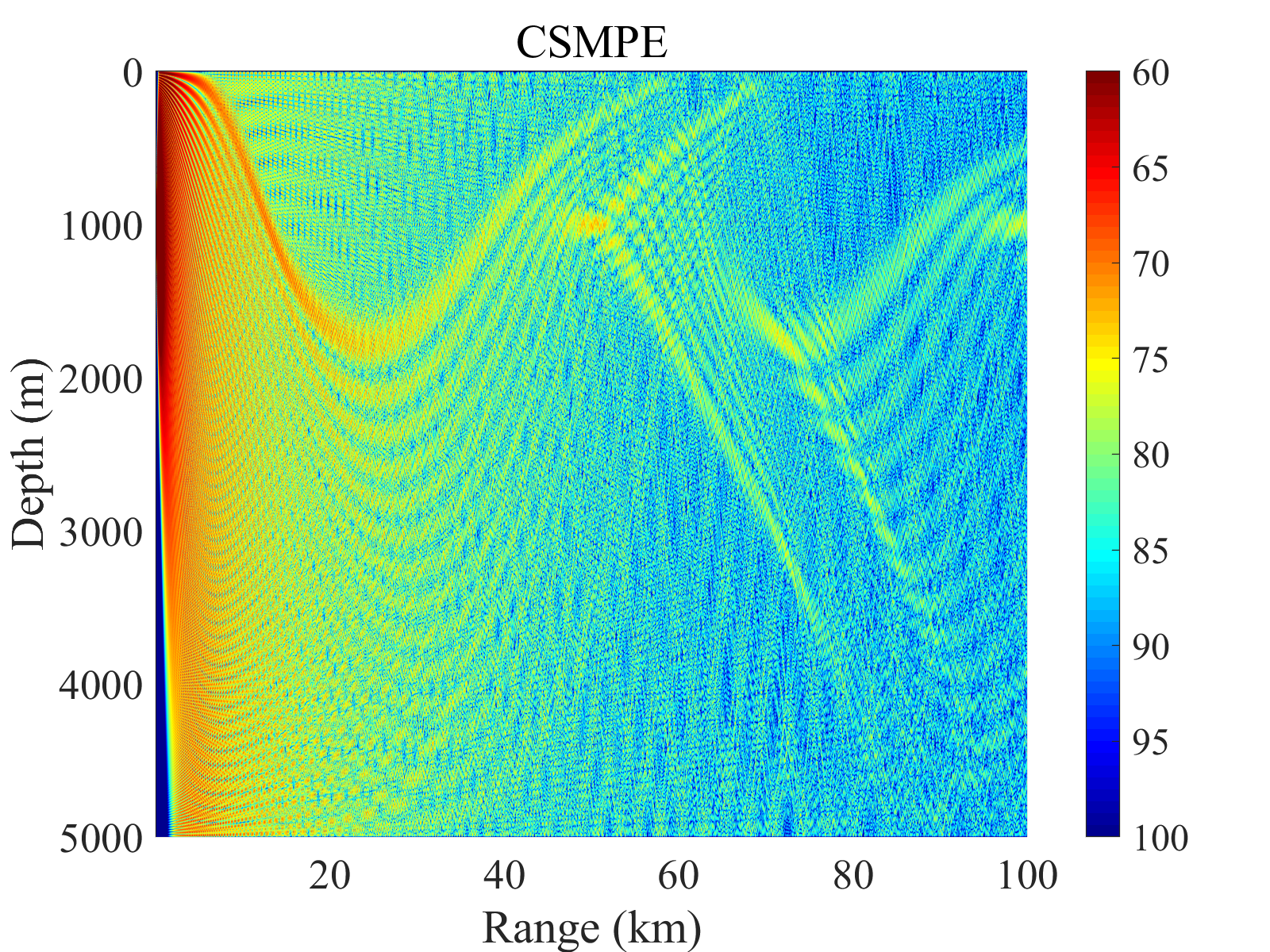}}
\caption{Acoustic fields of an ideal fluid waveguide with a Munk sound speed profile calculated using SCOOTER (a), KRAKEN (b), RAM (c) and CSMPE (d).\label{f5}}
\end{figure}

This case involves a deep-sea environment example. The marine environment in this case is intuitively shown in Fig.~\ref{f1}(b); the density of the seawater is uniform at $\rho(z)=1$ g/cm$^3$. The bottom of the sea is at a depth of $H=5000$ m. The sound source is located at a depth of $z_s=1000$ m, and the frequency is $f=50$ Hz. The sound speed is taken as the Munk sound speed profile \cite{Finn2011}, the general form of which is:
    \begin{equation}
        c(z)=1500.0\left\{1.0+0.0073\left[\frac{z-1300}{650}-1+\exp \left(-\frac{z-1300}{650}\right)\right]\right\}.
    \end{equation}
Since this example cannot provide an analytical solution, we can only compare the relative differences between the RAM and CSMPE results. We also calculate this example with SCOOTER and KRAKEN, and the results of both are used as references. Fig.~\ref{f5} clearly shows the spatial distributions of the TL calculated using the four programs. In the horizontal direction, $\Delta r$ in the four programs is taken as 20 m. In the vertical direction, the distance between discrete grid points in the SCOOTER, KRAKEN and RAM programs is taken as 1 m, and there are 5000 discrete points in total, i.e., $N_{RAM}=5000$; while the CSMPE takes 600 discrete points in the vertical direction; i.e., $N_{CSMPE}=600$. The phase velocity limits used by SCOOTER and KRAKEN are both 1500 to 6000 m/s, and KRAKEN has a total of 318 modes. The number of terms of the rational approximation is taken as $n=4$. As shown in Fig.~\ref{f5}, the TL fields of the four programs are very similar.

\begin{figure}[th]
\centerline{\includegraphics[width=\linewidth]{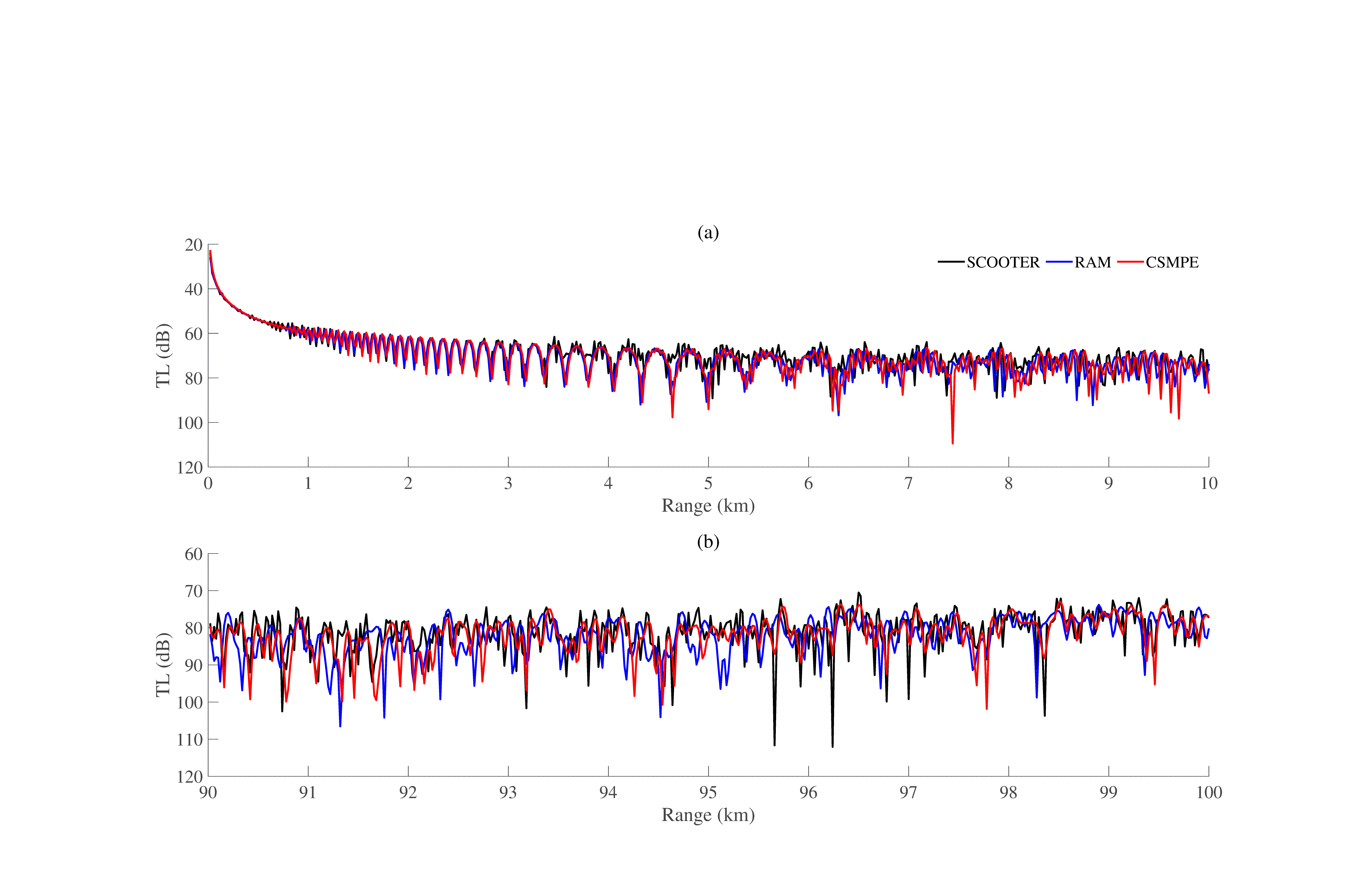}} \vspace*{8pt}
\caption{TL versus range calculated by SCOOTER, RAM and CSMPE at the depth $z_r=1000$ m.\label{f6}}
\end{figure}

To show the differences between the RAM and CSMPE results, Fig.~\ref{f6} shows the TL calculated by each of the two methods at the receiving depth $z_r=1000$ m. As shown in Fig.~\ref{f6}(a), in the near-field region, the RAM and CSMPE results are almost identical. When the horizontal range increases, the difference between the two programs gradually appears, and the TL curves no longer completely coincide, but the overall trends are identical. In the near field, the error between the two programs gradually increases with the range and becomes more scattered beyond 10 km. Fig.~\ref{f6}(b) shows the far-field TL at the receiving depth, showing that the TL of the programs has not only a magnitude error but also a phase error, which is quite normal for the PE model \cite{Finn2011,Yang2018}. Considering the entire field, the difference of most grid points is acceptable.

\subsection{Rough negative gradient sound speed profile in a shallow sea}
The sound speed profiles of the above two examples are smooth. Nevertheless, the CSMPE is still applicable for rough sound speed profiles. The sound speed profile in this case is intuitively shown in Fig.~\ref{f1}(c) and Table~\ref{tab1}; the density of the seawater is uniform at $\rho(z)=1$ g/cm$^3$. The bottom of the sea is at a depth of $H=400$ m. The sound source is located at a depth of $z_s=40$ m, the frequency of the sound source is $f=30$ Hz, and $n$ is taken as 8.
\begin{table}[th]
\tbl{\label{tab1}Sound speed profile in Fig.~\ref{f1}(c).}
{\begin{tabular}{@{}lccc@{}} \toprule
$z$ (m) & $c$ (m/s) & $z$ (m) & $c$ (m/s)\\ \colrule
0.00   & 1560.00 & 250.00  & 1520.00 \\
50.00  & 1555.00 & 300.00  & 1510.00 \\
100.00 & 1530.00 & 350.00  & 1505.00 \\
200.00 & 1525.00 & 400.00  & 1500.00 \\ \botrule
\end{tabular}}
\end{table}

\begin{figure}[th]
\centerline{\includegraphics[width=\linewidth]{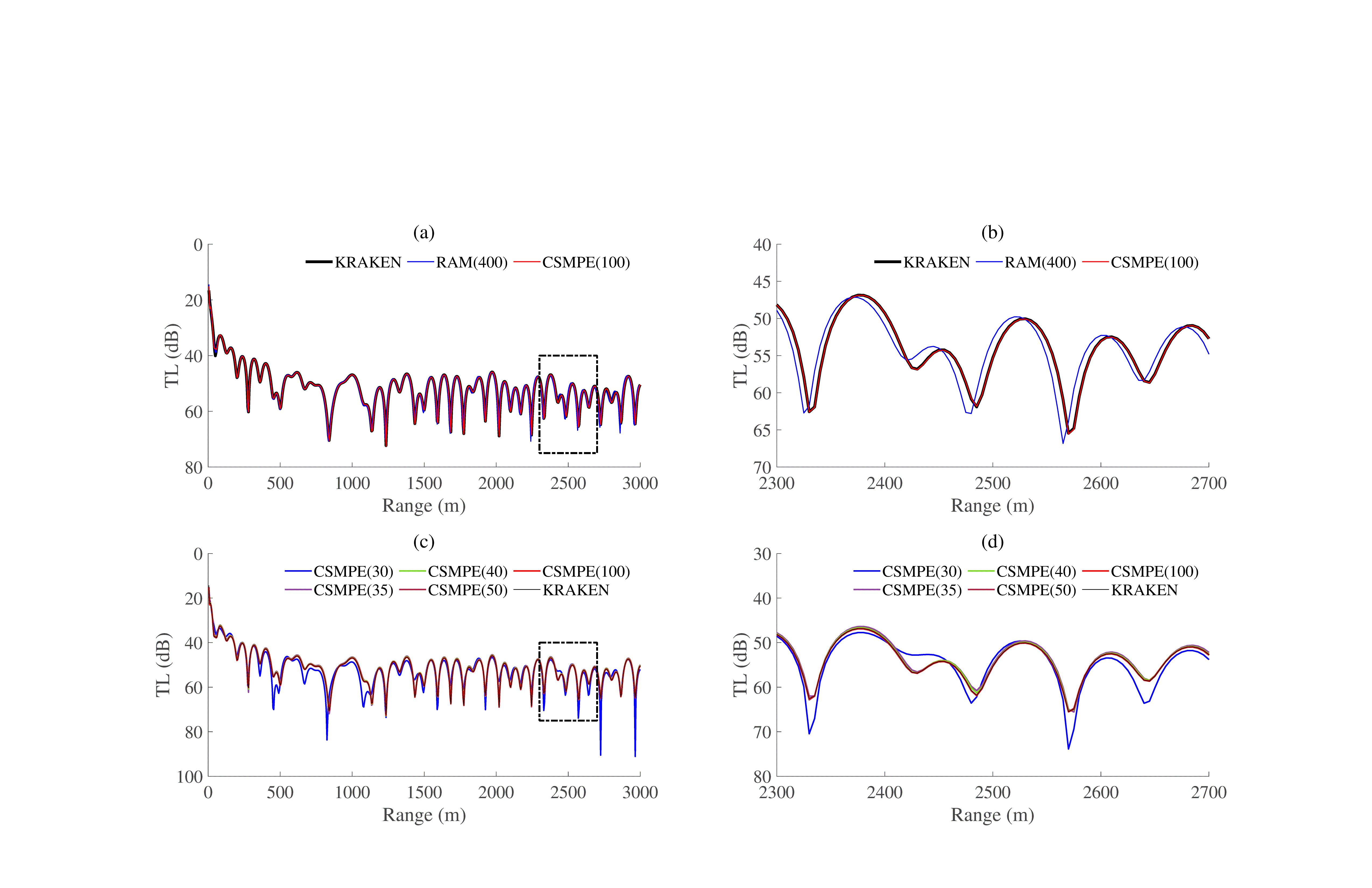}} \vspace*{8pt}
\caption{TL versus range calculated by the KRAKEN, RAM and CSMPE programs at the depth $z_r=40$ m (a); TL curves calculated by the CSMPE for different truncation orders $N_{CSMPE}$ (c); (b) and (d): magnified views of the dashed rectangles in (a) and (c), respectively. The numbers in brackets after RAM and CSMPE in the legend correspond to the numbers of discrete points taken in the vertical direction.\label{f7}}
\end{figure}

Figs.~\ref{f7}(a) and (b) show the variation in TL with range at a receiving depth of 40 m Fig.~\ref{f7}(a) indicates that the KRAKEN, RAM and CSMPE results are very similar and indistinguishable, which is sufficient to verify the correctness of the CSMPE program. Among them, the RAM uses 400 discrete points in the vertical direction, i.e., $N_{RAM}=400$, and the truncation order used by the CSMPE is $N_{CSMPE}=100$. Taking the KRAKEN results as a reference solution (the number of discrete points used by KRAKEN is also 400, the phase velocity limit is 1500 to 5500 m/s, with a total of 15 modes), Fig.~\ref{f7}(b) shows that the CSMPE results are closer to the KRAKEN results than the RAM results, although the number of discrete grid points used by the RAM is four times that used by the CSMPE. Figs.~\ref{f7}(c) and (d) illustrate the TL calculated by the CSMPE at a depth of 40 m under different truncation orders $N_{CSMPE}$. When $N_{CSMPE}$ increases from 30 to 100, the TL curve calculated by the CSMPE gradually coincides with that calculated by KRAKEN. Hence, for sound speed profiles that are not sufficiently smooth, the CSMPE can obtain a credible sound field by using a large truncation order $N_{CSMPE}$. This approach is important for measured sound speed profiles, although it will increase the runtime.

\section{Analysis of the runtime}
The CSMPE program developed in this paper and the RAM program use the same physical framework. The difference is the different discrete methods used for the depth operator $\mathcal{X}$. In this paper, the CSM is used to discretize the depth operator of Eq.~\eqref{eq:22}; therefore, we analyze only the difference in algorithmic complexity between the two programs. The processes of the forward Chebyshev transform and the backward Chebyshev transform are added to the CSMPE, and the algorithmic complexity of solving dense matrix equations is $O(N_{CSMPE}^3)$ in the ``split-step''. To reduce the complexity, we conduct preprocessing to perform LU decomposition of the dense matrix. Although the algorithmic complexity of LU decomposition is $O(N_{CSMPE}^3)$, this step is executed only once, and the algorithmic complexity of solving the linear equations in the ``split-step'' is reduced to $O(N_{CSMPE}^2)$. The depth operator in the RAM is discretized using the Galerkin method based on the FDM; the algorithmic complexity depends on the size $N_{RAM}$ of the tridiagonal matrix in the ``split-step''. Gaussian elimination involves sweeping downward to eliminate entries below the main diagonal followed by a back substitution sweeping upward \cite{Collins2006}. Hence, the algorithmic complexity of the RAM is $O(N_{RAM})$. Accordingly, the CSMPE has significantly greater complexity than the RAM, but the size of $N_{CSMPE}$ in the CSM is both the number of discrete points in the vertical direction and the truncated order of Chebyshev basis functions. Thus, the CSM can achieve higher accuracy with fewer discrete grid points. When $N_{RAM}\gg N_{CSMPE}$, the runtimes of the two programs must be tested by specific examples. To compare the speeds of the SM and the FDM, we test the runtimes of the above three examples, and the results are shown in Table \ref{tab2}. The listed runtimes in the table are the average of 10 test times. In the test, MATLAB 2019a is run on a Dell XPS 8930 desktop computer equipped with an Intel i7-8700K processor, and the memory is 16 GB.
\begin{table}[th]
\tbl{\label{tab2}Configurations and runtimes of the two examples.}
{\begin{tabular}{@{}cccccrcrc@{}} \toprule
Example & $f$ & $H$ & $n$ &$N_{RAM}$ &RAM&$N_{CSMPE}$&CSMPE & optimized CSMPE\\ \colrule
1 & 20 & 100  & 8 & 200  &  0.35 s & 25  &   0.17 s  &0.07 s\\
2 & 50 & 5000 & 4 & 5000 & 45.89 s & 600 & 191.39 s  &2.93 s\\
3 & 30 & 400  & 8 & 400  &  0.45 s & 100 &   0.95 s  &0.09 s\\ \botrule
\end{tabular}}
\end{table}
Table \ref{tab2} shows that in the three examples, the CSM has a longer runtime than the RAM despite having fewer discrete points. Thus, the CSMPE is slower than the RAM. Even considering that the RAM is a well-optimized code, the speed of the CSMPE is not satisfactory, which is its main disadvantage. 

Since this article considers the wide-angle PE model under the range-independent case, in this simple case, the CSMPE is still optimizable. In the process of solving Eq.~\eqref{eq:26}, we introduce the ``transfer matrix'' idea proposed by Xu et al. \cite{Xu2019}. Eq.~\eqref{eq:26} can be written as:
    \begin{equation}
        \label{eq:32}
        \prod_{j=1}^{n}\left(\mathbf{I}+\beta_{j,n}\mathbf{X}\right)\mathbf{\hat{p}}(r+\Delta r,x)=\exp{(ik_0\Delta r)}\prod_{j=1}^{n}\left(\mathbf{I}+\alpha_{j,n}\mathbf{X}\right)\mathbf{\hat{p}}(r,x).
    \end{equation}
Define the matrices as follows:
    \begin{equation}
        \label{eq:33}
        \mathbf{L}_j=\mathbf{I}+\beta_{j,n}\mathbf{X},\quad
        \mathbf{R}_j=\mathbf{I}+\alpha_{j,n}\mathbf{X}.
    \end{equation}
Let the transfer matrix $\mathbf{T}$ be as follows:
    \begin{equation}
    \label{eq:34}
        \mathbf{T}=\prod_{j=1}^{n}\left(\mathbf{L}_j^{-1}\mathbf{R}_j\right).
    \end{equation}
Thus, Eq.~\eqref{eq:26} is optimized as:
    \begin{equation}
    \label{eq:35}
        \mathbf{\hat{p}}(r+\Delta r,x)=\exp{(ik_0\Delta r)}\mathbf{T}\mathbf{\hat{p}}(r,x).
    \end{equation}
In this way, after the transfer matrix $\mathbf{T}$ is obtained, the ``split'' (solving $n$ linear equations) operation in each step in Eq.~\eqref{eq:26} is optimized to a matrix-vector multiplication operation. This will greatly reduce the amount of calculation in the CSMPE program. The last column of Table \ref{tab2} shows the running time of the CSMPE optimized using the transfer matrix idea. It can be seen from the table that the optimized CSMPE runs faster than the RAM. It should be emphasized that the optimized CSMPE is faster than the RAM because the CSMPE is only suitable for range-independent marine environments, which is equivalent to sacrificing the width of the function to improve the speed of the solution. This does not indicate that the spectral method is superior to the FDM in terms of computational speed. It is also worth mentioning that all three examples take the density as a constant, but the CSMPE can handle variations in the density with depth, which can be reflected in Eq.~\eqref{eq:25}.

\section{Conclusion and outlook}
The results of the numerical example with an analytical solution show that the CSMPE program can achieve higher computational accuracy with fewer discrete grid points than the RAM. When the CSM achieves approximately identical accuracy to the FDM, the number of grid points in the CSM is significantly smaller than that of the FDM. This finding is particularly useful for constructing a standard solution to the problem of underwater acoustic propagation over a large spatial area; in this problem, the use of the traditional FDM often requires too many spatial grids, while the CSM can effectively overcome this deficiency.

The comparison and analysis of these examples show that the CSM for the discrete PE model is feasible and reliable, the results are credible, and the CSM has higher accuracy than the classic PE model (the RAM program based on the FDM) in a range-independent environment. The disadvantage of the CSM is the large amount of calculations involved; in the calculation of the CSMPE program, it is necessary to solve the dense matrix equations multiple times, while the FDM must solve only the larger-scale tridiagonal matrix algebraic equations. However, for range-independent scenarios, the CSMPE can be optimized through the idea of the ``transfer matrix'' to greatly reduce the calculation burden to achieve a faster speed than the RAM.

In addition, the CSMPE program is used only in simple marine environments above a flat, horizontal ocean floor; that is, variations in the sound speed profile with range are not considered. In the future, we will attempt to generalize the CSM for use in complicated marine environments with terrain fluctuations, special sediments and sound speed profile changes with range based on the PE model. In its current state, the CSMPE is more suitable to provide high-precision reference standards for benchmark examples of the PE model in ocean acoustics.

\section*{Acknowledgments}
This work was supported by the National Key Research and Development Program of China (2016YFC1401800), the National Natural Science Foundation of China (61972406, 51709267) and the Project of National University of Defense Technology (4345161111L).

\end{document}